\documentclass[twocolumn,showpacs,preprintnumbers,amsmath,amssymb]{revtex4}

\usepackage{graphicx}
\usepackage{dcolumn}
\usepackage{bm}

\begin{document}

\preprint{APS/123-QED}

\title{Fourier-transfrom Ghost Imaging based on Compressive Sampling algorithm}

\author{Hui Wang}
\email{ami157@mail.siom.ac.cn}

\author{Shensheng Han}%
 \email{sshan@mail.shcnc.ac.cn}
\affiliation{Key Laboratory for Quantum Optics and Center for Cold Atom Physics of CAS,\\Shanghai Institution of Optics and Fine Mechanics,\\Chinese Academy of Sciences Shanghai 201800, China.}


\date{\today}

\begin{abstract}
A special algorithm for the Fourier-transform Ghost Imaging (GI)
scheme is discussed based on the Compressive Sampling (CS) theory.
The CS algorithm could also be used for the Fourier spectrum
reconstruction of pure phase object by setting a proper sensing
matrix. This could find its application in diffraction imaging of
X-ray, neutron and electron with higher efficiency and resolution.
Experiment results are also presented to prove the feasibility.
\end{abstract}

\maketitle

The diffraction imaging of X-ray, neutron and electron provide significant methods to reveal microstructure \cite{J.Miao, J.Miao2,I. K.,G. J.,B. Reuter,F. Pfeiffer,S. Marchesini}. Nevertheless, the traditional diffraction imaging by X-ray, neutron and electron in thermal state could only be used in periodic structure imaging. As for the aperiodic structure at the order of the source wavelength, according to the traditional wave theory, the coherent radiation source, such as Free Electron Laser (FEL), is essential for the imaging. However, since neutrons and electrons are fermions, their coherent sources with high brightness are in principle unavailable; although it's possible to obtain a coherent sources of photons with high brightness, the high requirement for brightness could only be met on the Synchrotron Radiation Facilities.

It has been proved that two-photon correlation Ghost Imaging (GI) could realize the diffraction imaging by taking advantages of thermal source \cite{Jing Cheng}, however, it is also limited by the long acquisition time, as the correlation theory calls for mass samples to guarantee the ensemble average. Actually, lots of methods were introduced to improve its Convergence \cite{M. Bache}, however, as long as it still takes the correlation algorithm, the concept of ensemble makes it less effective.

Recently the Compressive Sensing (CS) algorithm attracts more and more attention because of its extraordinary effect of reducing the samples \cite{Justin,E. J.,D. L.,E. J. Candes,J. Romberg,O. Katz}. GI and CS have similarities in extracting imaging: both are prepared with random sensing signal to ``express" the imaging and a ``point" detector to collect the result of expression. But they also have intrinsic difference: GI is based on accurate ``point by point" measure model \cite{Yanhua Shih,M. D¡¯Angelo}, while CS theory has proved that ``global random" measure model exhibits higher efficiency in imaging extraction \cite{Justin,E. J.,D. L.,E. J. Candes,J. Romberg,O. Katz}. Therefore, by combing GI and CS, it's possible to develop a brand-new imaging model with higher imaging efficiency and resolution.

At present, CS theory has already been introduced into GI scheme for real-space imaging\cite{O. Katz}. Since CS algorithm could be generalized to complex field, it in principle could also be used to Fourier GI scheme, where most spectrum information relies on the phase part of the field.

In view of these, we firstly propose a combination of the Fourier GI and CS in this letter. A recovery algorithm is investigated for the Fourier-Transform Ghost Imaging with Compressive Sampling (GICS) with experiment results. This algorithm could not only improve the imaging efficiency to reduce the radiation damage of the sample, but may also improve the spectrum resolution compared to the Correlation Ghost Imaging (CGI). Therefore GICS may provide a brand-new diffraction microscopic imaging technic of X-ray, neutron and electron for the aperiodic structure.

According to the CS theory, the algorithm is based on a corresponding relation between the imaging information $X$ and the detect signal $Y$ through a proper sensing matrix $A$ \cite{Justin,E. J.,D. L.,E. J. Candes,J. Romberg}:
\begin{eqnarray}\label{eq1}
 AX=Y
 \end{eqnarray}

If we sample K times, A should be a $K\times N$ matrix, and $Y$ should be a known K-element vector. Then by solving a convex optimization program, it gives an optimal result of $X$ with N-element in the same expressing space as A\cite{Justin,E. J.,D. L.,E. J. Candes,J. Romberg}.

\begin{figure}
\includegraphics[width=8.5cm]{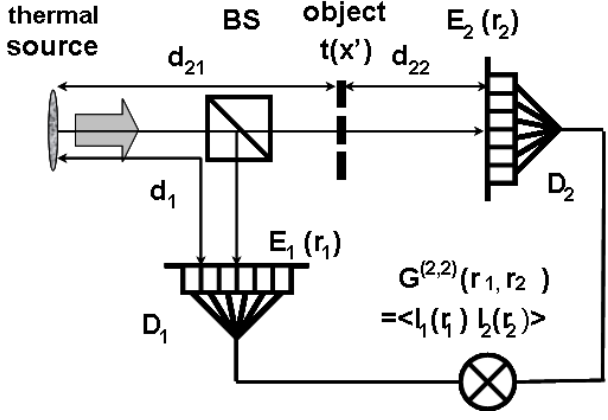}
\caption{The lensless Fourier-transform GI scheme. Beam splitter (BS) is introduced to divide the field into two arms: in the test arm, field propagates freely $d_{21}$ to an object and then $d_{22}$ to array detector $D_2$; in the reference arm, field propagates d1 to another array detector $D_1$.$d_1 = d_{21} + d_{22}$.}
\end{figure}

In our case, the lensless Fourier-transform GI scheme is shown in Fig 1 \cite{Jing Cheng}. The thermal field from the source S is divided by the beam splitter (BS) into test and reference arms. To perform the Fourier-transform GI, there should be $d_1 = d_{21} + d_{22}$ for the scheme.
Based on this scheme, we could express the intensity $I_w(r_2)$ on the test detector D2 as:

\begin{eqnarray}\label{eq3}
 I_w (r_2 ) &\propto& \int\limits_{obj} {dxdx'E^* (x)E(x')} t^* (x)t(x')\\ \nonumber
            & &\times  \exp \left\{ {\frac{{i\pi }}{{\lambda d_{22} }}\left[ {(x - r_2 )^2  - (x' - r_2 )^2 } \right]} \right\}
 \end{eqnarray}

Where the integration of $x$ and $x'$ is over the object plane. By comparing with the CS theory, the object information is located on the right side of (\ref{eq3}), thus $I_w(r_2)$ could be considered as the known detect signal $Y$. Since our imaging goal is the Fourier-transform of $t(x)$, to process the CS algorithm, it's necessary to establish a corresponding relation like (\ref{eq1}), and the sensing matrix $A$ must be related to the signal $I_r(r_1)$ from the reference detector $D_1$, because the spectrum used to be expected to show up on the array detector $D_1(r_1)$.

If we consider the array of $D_1$ to be large enough, then (\ref{eq3}) could be rewritten approximately as
\begin{widetext}
\begin{eqnarray}\label{eq4}
 I_w (r_2 ) &\propto& \int\limits_{ref} {dr_1 dr_1 'E^* (r_1 )E(r_1 ')} \int\limits_{obj} {dxdx'\exp \left\{ { - \frac{{i\pi }}{{\lambda d_{22} }}\left[ {(x - r_1 )^2  - (x' - r_1 ')^2 } \right]} \right\}}\\ \nonumber
            & & \times t^* (x)t(x')\exp \left\{ {\frac{{i\pi }}{{\lambda d_{22} }}\left[ {(x - r_2 )^2  - (x' - r_2 )^2 } \right]} \right\} \\ \nonumber
            & = & \int\limits_{ref} {dr_1 dr_1 'E^* (r_1 )E(r_1 ')} \exp \left\{ { - \frac{{i\pi }}{{\lambda d_{22} }}\left( {r_1 ^2  - r_1 '^2 } \right)} \right\} \int\limits_{obj} {dx dx'} t^* (x)t(x')\exp \left\{ {\frac{{i2\pi }}{{\lambda d_{22} }}\left[ {(r_1  - r_2 )x - (r_1 ' - r_2 )x'} \right]} \right\} \\ \nonumber
            & = &\int\limits_{ref} {dr_1 dr_1 'E^* (r_1 )E(r_1 ')\exp \left\{ { - \frac{{i\pi }}{{\lambda d_{22} }}\left( {r_1 ^2  - r_1 '^2 } \right)} \right\}} T^* (f_1  = \frac{{r_1  - r_2 }}{{\lambda d_{22} }})T(f_1 ' = \frac{{r_1 ' - r_2 }}{{\lambda d_{22} }})
 \end{eqnarray}
\end{widetext}

Compare (\ref{eq4}) with (\ref{eq1}), the sensing matrix A could then be expressed as $E^* (r_1 )E(r_1 ')\exp \left\{ { - {\raise0.7ex\hbox{${i\pi }$} \!\mathord{\left/ {\vphantom {{i\pi } {\lambda d_{22} }}}\right.\kern-\nulldelimiterspace}\!\lower0.7ex\hbox{${\lambda d_{22} }$}}\left( {r_1 ^2  - r_1 '^2 } \right)} \right\}$.

Worth to note, unlike the sensing matrix $A$ introduced in the bucket real-space GI scheme \cite{O. Katz}, the matrix $A$ here is a $K\times N^2$ one rather than a $K\times N$ one. However£¬since the reference detector $D1$ could only record N intensities, the $N^2-N$ interference terms of $A$ when $r_1  \ne r_1 '$ are in general unobtainable .

Generally, the only way available to get the exact field distribution on $D_1$ in GI scheme is by calculating for the controllable source scheme or by introducing another known reference field $E'$ for the homodyne detection. However, both are unavailable for the diffraction imaging of X-ray, neutron and electron.

 Notice if we consider each row of $A$ to be a $N\times N$ matrix $A_{sub}$ for each sample, the final imaging spectrum result obtained from CGI, $\left| {T\left( f \right)} \right|$, should lie just on the diagonal line of the CS solution $X$ (also a $N\times N$ matrix), corresponding to the known intensity distribution of $D_1$ on the diagonal line of $A_{sub}$. According to the CS algorithm, the relative distribution of $\left| {T\left( f \right)} \right|$ could be expressed through the known $I_r(r_1)$ of K sampling, while the existence of other non-diagonal elements just make the relation (\ref{eq4}) hold. Therefore, if we only want to get $\left| {T\left( f \right)} \right|$, the accurate position of the non-diagonal elements, whose phases distribute randomly, become less important.

Concretely speaking, if we properly conjecture the phase distribution on $D_1$ to be $\Phi \left( r_1 \right)$ for each sample based on the concrete scheme, then each row of $A$ could be written as
\begin{eqnarray}\label{eq5}
 A(r_1 ,r_1 ') &=& \sqrt {I_r (r_1 )I_r (r_1 ')} \exp \left\{ { - i\left[ {\Phi \left( {r_1 } \right) - \Phi \left( {r_1 '} \right)} \right]} \right\}\\ \nonumber
               & & \times \exp \left\{ { - \frac{{i\pi }}{{\lambda d_{22} }}\left( {r_1^2  - r_1 '^2 } \right)} \right\}
 \end{eqnarray}

Although the randomly conjectured phase distribution $[\Phi \left( {r_1 } \right) - \Phi \left( {r_1'} \right)]$ deviates from the true value, the intensity distribution $\sqrt {I_r (r_1 )I_r (r_1 ')}$ is random too, which makes the non-diagonal elements $\sqrt {I_r (r_1 )I_r (r_1 ')} \exp \left\{ { - i\left[ {\Phi \left( {r_1 } \right) - \Phi \left( {r_1 '} \right)} \right]} \right\}$ in whole randomly distributed. When the number of conjectured values is large enough, the values, though random, tend to cover all the true value of non-diagonal elements and establish $A_{sub}$. Since we don't care about $T^* (f_1)T(f_1 ')$, then follow the normal CS algorithm we could extract $\left| {T\left( f \right)} \right|$ along the diagonal line.

Obviously, (\ref{eq4}) is not a strict equation when the conjectured values aren't enough, just as a non-ideal bucket detector for the real-space imaging \cite{O. Katz}, whose influence to the sensing matrix is ignored. However, since we always keep the true intensities $I_r (r_1 )$ on the diagonal line of $A_{sub}$, this approximation has more influence on the efficiency to extract the information rather than destroying the expressing basis for $\left| {T\left( f \right)} \right|$. Also, there's little relation in applications with strict equation for the CS algorithm to perform, thus it makes little sense to pursue the strict condition for CS algorithm. On the contrary, it's proper to develop the CS algorithm adaptive to relations with different approximations. And still, GICS, as a brand new imaging algorithm instead of CGI, brings improvement of the low efficiency and limit resolution to the traditional CGI.

Our experiment scheme is the same as shown in Fig.1, with source size $\sigma$ set to be 3 mm and wavelength is $\lambda=0.532\mu m$, $d_{21}=20 cm, d_{22}=5 cm, d_1=d_{21}+d_{22}$, $D_1$ is an array detector and $D_2$ is a point detector. The object is a one dimension pure phase five-slit, whose transmission function and spectrum intensity $\left| {T\left( f \right)} \right|$ is shown in FIG.2. The laser, CGI and GICS spectrum results of $\left| {T\left( f \right)} \right|$ are shown in FIG.3 for comparison, where the GICS algorithm is based on the Spectral Projected Gradient for L1 minimization (SPGL1) to solve the Basis pursuit (BP) problem in (\ref{eq1}) and we both take 100 off-diagonal elements of $A_{sub}$ conjectured in FIG.3(c) and (e). Obviously, from the same sample data GICS (shown in FIG.3 (c) and (e)) obtains a much better result than traditional CGI result (shown in FIG.3 (b) and (d)), which means improved extract efficiency.

\begin{figure}
\includegraphics[width=8.5cm]{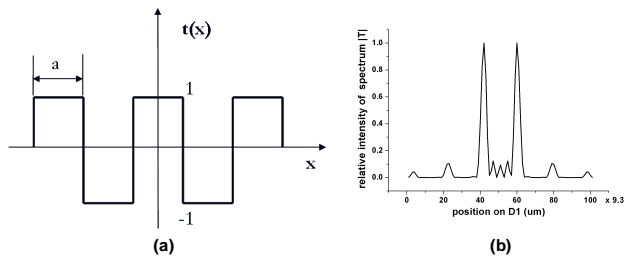}
\caption{Transmission function and spectrum intensity $\left| {T\left( f \right)} \right|$ of the pure phase five-slit object. (a)Transmission function of the object, the slit wide $a$ is 150 $\mu$m.(b)Spectrum intensity $\left| {T\left( f \right)} \right|$ of the object observed from a lens of 5 cm focal length.}
\end{figure}

\begin{figure}
\includegraphics[width=8.5cm]{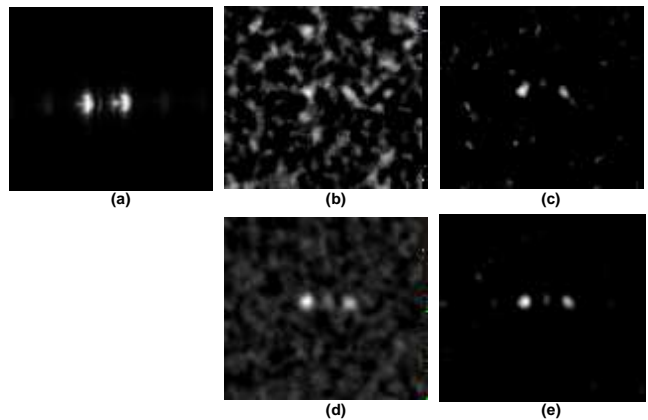}
\caption{The experiment reconstruction results of spectrum intensity $\left| {T\left( f \right)} \right|$ comparison among laser, CGI and GICS. (a) Reconstruction results by laser illumination; (b) Reconstruction results by traditional CGI, the number of sample $K=100$;(c) Reconstruction results by GICS (SPGL1), the number of sample $K=100$ and number of iterations $I=30$; (d) Reconstruction results by traditional CGI, $K=1000$;(e) Reconstruction results by GICS (SPGL1), $K=1000$ and $I=30$;}
\end{figure}


Besides the higher extract efficiency, GICS could also introduce super-resolution beyond CGI. This is of great importance to the reconstruction of object, especially for the Phase retrieval of nonperiodic objects\cite{J.Miao}, where oversampling method is available. It can be proved that the resolution of lensless Fourier-transform CGI is limited to the transversal coherent length on the reference detector $D_2$. However, by taking advantage of GICS the limitation could be broken.

The experiment for super-resolution is shown in FIG.4. The scheme parameter is the same as FIG.3, only with the object changed to a double-slit. The slit width is $200\mu m$, and the slit separation is $600\mu m$. The transversal coherent length on $D_2$ is calculated by $l_c=\lambda d_1\diagup \sigma\approx44\mu m$, while the peak separation of the spectrum $\left| {T\left( f \right)} \right|$ is $32 \mu m$. Therefore the traditional CGI(shown in FIG.4 (b)) could not resolve the spectrum, and GICS(shown in FIG.4 (c)) could realize super-resolution. The GICS result in FIG.4 (c) is also based on BP problem in SPGL1 but without any off-diagonal element of $A_{sub}$ conjectured. This is also a proof for the role of the off-diagonal element discussed before, since it's still possible to extract $\left| {T\left( f \right)} \right|$ only based on the on-diagonal elements or the intensities on $D_2$. In this experiment, both FIG.4 (b) and (c) have been performed through spatial averaging technique.

\begin{figure}
\includegraphics[width=8.5cm]{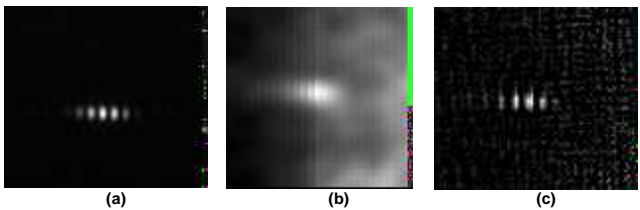}
\caption{Spectrum super-resolution result of GICS. (a) Spectrum result by laser illumination; (b) Spectrum result by traditional CGI with sample number to be $K=1000$; (c) Spectrum result by GICS (BP problem in SPGL1), where $K=1000$ and $I=500$.}
\end{figure}

The improvement of extract efficiency and resolution introduced by GICS is also tenable for complicated spectrum. FIG.5 shows the GICS spectrum result of a complicated object ``zhong" ring, compared with the results of laser illumination and CGI. The experiment scheme is also the same as in FIG.3,and the diameter of the object ring is $0.8 mm$. Still the GICS result in FIG.5 (d) is based on BP problem in SPGL1 without off-diagonal elements conjectured. and both CGI and GICS results (shown in FIG.5 (b) and (c)) have been performed through spatial averaging technique.

\begin{figure}
\includegraphics[width=8.5cm]{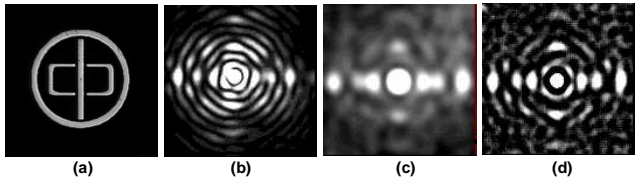}
\caption{Spectrum result of a complicated object ``zhong" ring. (a) Amplitude distribution of the complicated object "zhong" ring; (b) Spectrum result by laser illumination; (d) Spectrum result by CCS, where the sample number is $K=4000$; (d) Spectrum result by GICS (BP problem in SPGL1), where sample number is $K=4000$ and iteration number is $I=100$.}
\end{figure}

In conclusion, A new algorithm based on the lensless Fourier-transform GICS is presented in this letter. This scheme opens up a new way for the diffraction imaging of X-ray, neutron and electron with higher efficiency and super-resolution. The GICS, however, is much more than a algorithm revolution. From the point of microscopic essence, quantum fluctuations of field are generally described via high order correlation, which is based on accurate classical ``point by point" measurement: the traditional imaging technic is first order correlation (single-photon) system, while CGI is a second order correlation (two-photon) system. CS, on the contrary, stems from a ``global random" sensing measurement. By introducing CS algorithm into GI, it also initiates a new discussion on its microscopic quantum mechanism. To find a reasonable explanation for the microscopic essence of GICS is a challenging problem, which will help to develop new imaging model with higher efficiency.

The work is supported by the Hi-Tech Research
and Development Program of China under Grant No.
2006AA12Z115, National Natural Science Foundation of China under Grant Project No.
60877009, and Shanghai Natural Science Foundation under Grant
Project No. 09JC1415000.


\begin{thebibliography}{99}
\bibitem{J.Miao}J.Miao, D.Sayre, and H. N. Chapman, J. Opt.Soc.Am A 15, 1662 (1998).
\bibitem{J.Miao2}J.Miao, P. Charalambous, J. Kirz, and D. Sayre, Nature(London) 400, 342 (1999).
\bibitem{I. K.}I. K. Robinson, I. A. Vartanyants, G. J. Williams, M.A.Pfeifer, and J. A. Pitney, Phys. Rev. Lett. 87, 195505 (2001)
\bibitem{G. J.}G. J. Williams, M. A. Pfeifer, I. A.Vartanyants, and I. K. Robinson, Phys. Rev. Lett. 90,175501 (2003).
\bibitem{B. Reuter}B. Reuter and H. Mahr, Experiments with Fourier transform holograms using 4.49nm x-rays, J. Phys. E., Vol.9(9): 746-751, (1976).
\bibitem{F. Pfeiffer}F. Pfeiffer, T. Weitkamp, O. Bunk and C. David, Nature Physics, 2, 258, (2006).
\bibitem{S. Marchesini}S. Marchesini, H. He, H.N. Chapman, S.P. Hau-Riege, A. Noy, M.R. Howells, U. Weierstall, and J.C.H. Spence, Phys. Rev. B 68, 140101(R) (2003).
\bibitem{Jing Cheng}Jing Cheng, and Shensheng Han. Phys.Rev. Lett. 92, 9 (2004).
\bibitem{M. Bache}M. Bache, E. Brambilla, A. Gatti and L.A. Lugiato, Opt. Express 12, 24(2004)
\bibitem{Justin} Justin Romberg, IEEE SIGNAL PROC MAG. March, 14-20(2008)
\bibitem{E. J.} E. J. Candes and M. B. Wakin, IEEE Sig. Proc. Mag. March, 21-30 (2008).
\bibitem{D. L.}D. L. Donoho and Y. Tsaig, IEEE Trans. Inform. Theory. 54, 4789-4812 (2006).
\bibitem{E. J. Candes}E. J. Candes, J. Romberg, and T. Tao, IEEE Trans. Inform. Theory, 52, 5406-5425 (2006).
\bibitem{J. Romberg}J. Romberg, IEEE Sig. Proc. Mag. March, 14 (2008).
\bibitem{O. Katz}O. Katz, Y. Bromberg, and Y. Silberberg, Appl. phys. Lett. 95, 131110 (2009).
\bibitem{Yanhua Shih}Yanhua Shih, Front. Phys. China, 2(2): 125¨D152, (2007)
\bibitem{M. D¡¯Angelo}M. D'Angelo and Y.H. Shih, Laser Phys. Lett. 2, 12: 567¨C596 (2005)



\end{thebibliography}
\end{document}